# Dual Gate Graphene FETs with $f_T$ of 50 GHz

Yu-Ming Lin, *Member, IEEE,* Hsin-Ying Chiu, Keith A. Jenkins, *Senior Member, IEEE,*

Damon B. Farmer, Phaedon Avouris, *Member, IEEE*, Alberto Valdes-Garcia, *Member, IEEE*

**Abstract**

A dual-gate graphene field-effect transistors is presented, which shows improved RF performance by reducing the access resistance using electrostatic doping. With a carrier mobility of 2700 cm$^2$/Vs, a cutoff frequency of 50 GHz is demonstrated in a 350-nm gate length device. This $f_T$ value is the highest frequency reported to date for any graphene transistor, and it also exceeds that of Si MOSFETs at the same gate length, illustrating the potential of graphene for RF applications.

This work is supported by DARPA under contract FA8650-08-C-7838 through the CERA program.

The authors are with IBM T. J. Watson Research Center, Yorktown Heights, NY 10598 (email: yming@us.ibm.com)

The views, opinions, and/or findings contained in the article are those of the authors and should not be interpreted as representing the official views of the DARPA or the Department of Defense.

I. INTRODUCTION

Graphene is a monolayer of carbon atoms in a honeycomb lattice, which has attracted considerable attention over the last few years due to its unique electronic properties [1]. With a high saturation velocity ($5.5 \times 10^7$ cm/s) [2], graphene is considered a very promising candidate for mmWave (millimeter wave) applications. In addition, the ultra-thin body thickness of graphene offers ideal two-dimensional electrostatics for the ultimately scaled-down device. Recently, cutoff frequencies in the GHz regime have been demonstrated in top-gated graphene transistors built on exfoliated single-layer graphene sheets [3,4] and few-layer graphene grown on SiC substrates [5]. It was found that, in top-gated graphene FETs, the RF performance was primarily limited by the mobility degradation of graphene after oxide deposition [3,6]. Recently, Kim et al. have reported progress in enhancing the carrier mobility in top-gated graphene FETs by using a thin layer of naturally oxidized $Al_2O_3$ as the nucleation layer for the deposition of high-k gate dielectrics [7]. Another important factor that strongly affects the overall RF performance of graphene transistors is the access resistance between the source/drain contacts and the gated graphene channel. The access regions are required to minimize the capacitance between the gate and source/drain electrodes in a top-gated FET structure. However, since this access region consists of only a monolayer of un-gated graphene, the sheet resistance is much higher than that of heavily doped Si used in conventional MOSFETs, and is comparable to the resistance of the gated graphene channel. Access resistance becomes particularly important when the gate length shrinks, and may hinder further improvement of RF performance with the down-scaling of gate length in graphene transistors.

This letter describes the use of a dual-gate graphene field-effect transistor to investigate the impact of the access resistance on the RF performance. In this structure, the access resistance of the graphene transistor is modulated by the back gate through electrostatic doping. By varying the back gate voltage, the access resistance is reduced by more than half, leading to a four-fold increase of transconductance in a 350-nm-gate graphene FET. Combined with a carrier mobility of 2700 $cm^2$/Vs enabled by an improved oxide deposition process, a cutoff frequency of 50 GHz in the dual-gate graphene FET is achieved.

## II. DEVICE FABRICATION

The device structure of the dual-gate graphene FET (GFET) is shown in Fig 1(a). Single-layer graphene was deposited by mechanical exfoliation on high-resistivity Si substrates (>10 kΩ·cm) covered with 300-nm-thick thermal oxide. The source and drain electrodes made of Pd/Au metals (20nm/40nm thick) were fabricated by e-beam lithography and lift-off. The oxide deposition process described in Ref. [7] was adopted here to form a layer of 12-nm-thick $Al_2O_3$ by ALD (atomic layer deposition) as the top-gate dielectric. Fig. 1(c) shows that the device transfer characteristics were not appreciably degraded after this dielectric process. Finally, the Pd/Au (20 nm/40 nm) metal stack was deposited as the top-gate electrode. Fig. 1(b) shows the SEM image of the double-channel graphene transistor with a gate length of $L_g$ = 350 nm. The width of each channel is 27 μm and the spacing between the top-gate electrode and the source/drain contacts is 300 nm.

## III. RESULTS AND DISCUSSION

The Si substrate is used as the global back gate while the top gate serves as the main gate terminal for regular FET operations. Fig. 2(a) shows the transfer characteristics of the graphene FET at different back-gate voltages $V_{BG}$. At $V_{BG}$ = 0 V, the graphene FET exhibits ambipolar transfer characteristics with a current minimum at $V_{TG}$ = 0.7V. This ambipolar transport reflects the gapless nature of graphene band structure. The current minimum corresponds to the Dirac point, where the total carrier density of electrons and holes in the graphene channel becomes minimal. The Dirac voltage $V_{DRC}$, defined as the top-gate voltage at the Dirac point, is linearly dependent on $V_{BG}$, as shown in Fig. 2(b), and the slope ($\Delta V_{BG}/\Delta V_{DRC} \cong 35$) can be used to determine the capacitance $C_{TG}$ of the top-gate dielectrics. Using the back-gate capacitance value of $C_{BG}$ = 11.6 nF/cm$^2$, $C_{TG}$ is estimated to be 0.40 μF/cm$^2$.

In Fig. 2(a), the current at the Dirac point decreases with increasing $V_{BG}$, indicating an increasing series resistance with $V_{BG}$. This is because the back-gate not only shifts the threshold voltage of the graphene transistor, it also modulates the graphene not covered by the top gate. This additional resistance is analogous to the access resistance in conventional Si MOSFETs. The total resistance of the graphene device $R_T$ is modeled as the sum of an ideal graphene channel modulated by the top gate and a series resistance $R_S$ [2,7], and $R_T$ is given by

$$R_T = R_S + \left[ e\mu \frac{W}{L_G} \sqrt{n_0^2 + [\frac{C_{tot}}{e} \cdot (V_{TG} - V_{DRC})]^2} \right]^{-1}, \quad (1)$$

where $\mu$ is the field-effect mobility and $n_0$ is the minimum sheet carrier density determined by disorder and thermal excitation. $C_{tot}$ is the total top-gate capacitance given by $C_{tot}^{-1} = C_{TG}^{-1} + C_Q^{-1}$, where $C_Q$ is the quantum capacitance of graphene. For a relevant carrier density of $\sim 5\times10^{12}$ cm$^{-2}$, $C_Q \cong 3\mu F/cm^2$. Based on Eq. (1), a minimum carrier density $n_0 \sim 5\times10^{11}$ cm$^{-2}$ and a field-effect mobility $\mu \sim 2700$ cm$^2$/Vs, both independent of $V_{BG}$, was obtained for the graphene channel using the extraction method described in Ref. [2,7]. In the following, we show that while the mobility remains constant as $V_{BG}$ varies, the device performance can be improved by optimizing other device parameters. Fig. 2(b) shows the extracted series resistance $R_S$ of the graphene device as a function of $V_{BG}$, where $R_S$ rises with increasing $V_{BG}$ up to 40 V. This $V_{BG}$ gate dependence of $R_S$ due to un-gated graphene is consistent with the trend shown in Fig. 1(c). It is noted that that $R_S$ also includes the contact resistance between graphene and metal electrodes. To achieve the optimal RF performance in dual-gate graphene transistors, it is necessary to bias the $V_{BG}$ properly so that $R_S$ is at its minimum.

Fig. 2(c) shows the transconductance $g_m$ of the graphene device at different back-gate voltages. The impact of the series resistance on the device performance is evident, as can be seen in the increasing peak values of the p-branch $g_m$ when $V_{BG}$ decreases from 40 V to -40V. At $V_{BG}$ = -40V and $V_{TG}$ = 1.5, the graphene FET reaches a peak $g_m$ of -0.22 mS/μm. It is noted that while the series resistance is reduced by roughly half from $V_{BG}$= 40 V to -40 V, the peak p-type $g_m$ is enhanced by four times, changing from -0.05 to -0.22 mS/μm. The output characteristics at $V_{BG}$ = -40V are shown in Fig. 2(d).

To assess the RF characteristics of the graphene FET, on-chip microwave measurements were carried out up to 30 GHz. The measured S-parameters were de-embedded using specific "short" and "open" structures with identical layouts, excluding the graphene channel, to remove the effects of the parasitic capacitance and the resistance associated with the pads and connections. The use of high-resistivity Si substrates allows for a dc back-gate bias, while at the same time enabling RF operation without significant signal loss. Based on the results in Fig. 2(c), $V_{BG}$ was kept at -40V in order to achieve the highest RF performance in the dual-gate graphene FET. Fig. 3(a) shows the current gain $|h_{21}|$ from the measured S-parameters at $V_{TG}$ = 1.6 V and a drain bias $V_{DS}$ = 0.8 V, yielding a cut-off frequency $f_T$ of 50 GHz. The de-embedded current gain $|h_{21}|$ exhibits the -20dB/dec frequency dependence as expected for a well-behaved FET. This $f_T$ value is the highest frequency reported to date for any graphene transistor, and it also exceeds that of Si MOSFETs (~25 GHz) at the same gate length of 350 nm [8]. Fig. 3(b) shows the peak $g_m$ of p-type graphene FETs as a function of the series resistance modulated by $V_{BG}$. The well-known relation $f_T = g_m/(2\pi C)$ established for conventional FETs has recently been demonstrated to be also valid for graphene devices [3]. Based on the measured $f_T$ of 50 GHz for $g_m$=0.22 mS, the right axis of Fig. 3(b) shows the projected $f_T$ as a function of $R_S$, illustrating a four-fold improvement in $f_T$ as $R_S$ decreases from 110Ω to 50Ω.

## IV. CONCLUSION

A dual-gate graphene transistor is fabricated, showing improved RF performance by optimizing the series resistance. A cutoff frequency of 50 GHz is demonstrated for a gate length of 350 nm. This value exceeds that of Si MOSFETs at the same gate length, illustrating the potential of graphene for RF applications. In addition, while a global back gate is used here to optimize the access resistance in a dual-get graphene FET, the results can be generalized to other graphene FET structures. It is expected that similar performance enhancement can be achieved through other techniques such as local bottom gates or selective doping [9] to modulate the resistance in the access regions.

ACKNOWLEDGMENT

The authors would like to thank C. Y. Sung and F. Xia for insightful discussions, and B. Ek and J. Bucchignano for technical assistance. They also thank E. Tutuc and S. Kim for the discussions on oxide deposition.

FIGURES

Fig. 1 (a) Device schematic of the dual-gate graphene transistor. (b) SEM image of a dual-channel graphene transistor. The channel width is 27 μm and the gate length is 350 nm for each channel. (c) Measured channel conductance as function of back-gate voltage of a graphane device before and after the deposition of 12-nm-thick ALD $Al_2O_3$. Prior to the ALD process, a layer of 2-nm Al is deposited and oxidized as the nucleation layer.

Fig. 2 (a) Transfer characteristics of the GFET at various back-gate voltages and $V_D$ = 0.8 V. (b) Series resistance of the GFET and the Dirac voltage $V_{DRC}$ as a function of back-gate voltage. (c) Transconductance of the GFET at various back-gate voltages. The drain bias is 0.8V. (d) Output characteristics at $V_{BG}$ = -40V.

Fig. 3 (a) RF performance of a 350-nm-gate GFET, showing a current gain at -20dB/dec and a cut-off frequency $f_T$ of 50GHz. (b) Peak transconductance as a function of series resistance $R_S$ derived from Figs. 2(b) and (c). The projected cut-off frequency $f_T = g_m/2\pi C$ is shown on the right axis.

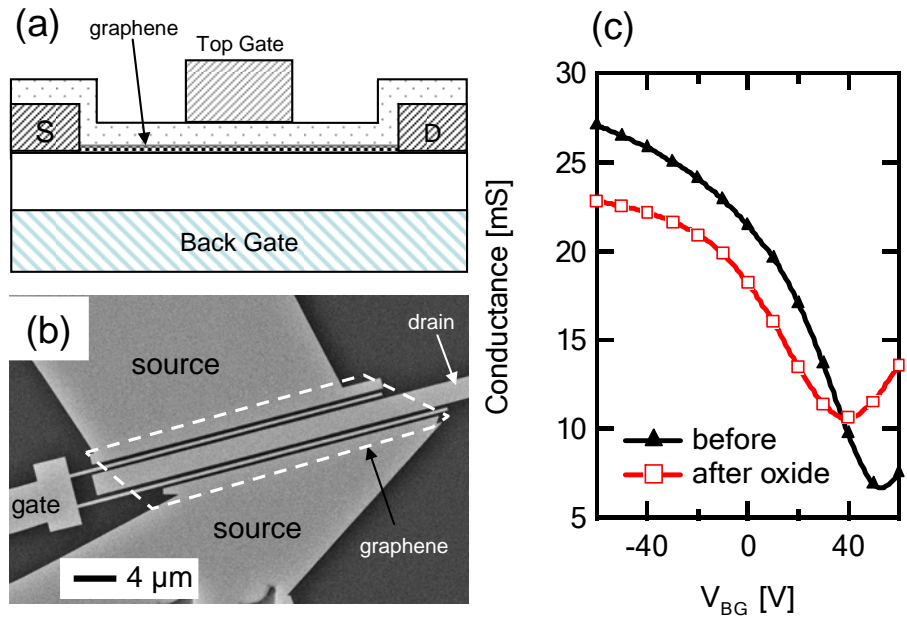

Fig. 1

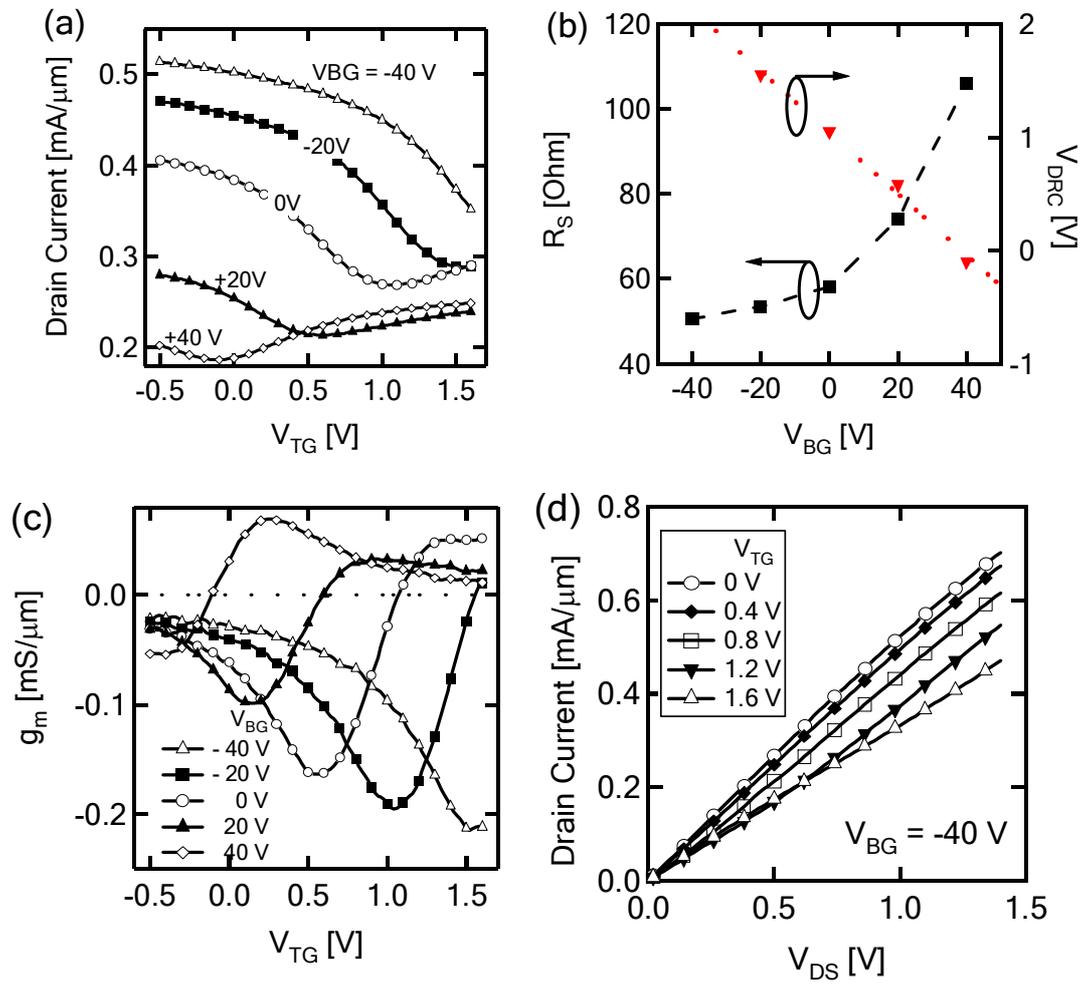

Fig. 2

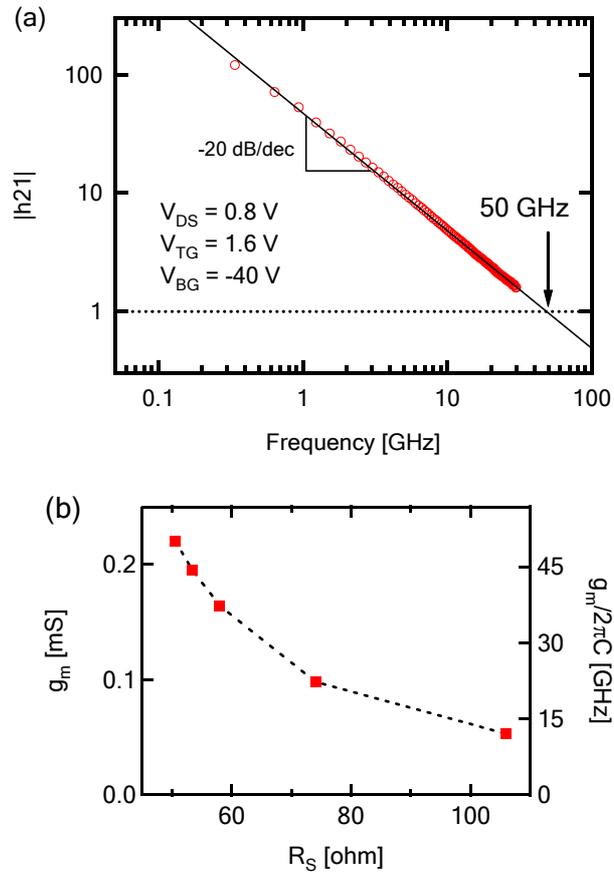

Fig. 3